\documentclass[conference]{IEEEtran}
\IEEEoverridecommandlockouts
\usepackage{lipsum}
\usepackage{cite}
\usepackage{balance}
\usepackage{amsmath,amssymb,amsfonts}
\usepackage{graphicx}
\usepackage{textcomp}
\usepackage{xcolor}
\usepackage[utf8]{inputenc}
\usepackage[T1]{fontenc}
\usepackage{hyperref} 
\usepackage{algorithm}
\usepackage{algorithmic}
\usepackage{comment}
\def\BibTeX{{\rm B\kern-.05em{\sc i\kern-.025em b}\kern-.08em
    T\kern-.1667em\lower.7ex\hbox{E}\kern-.125emX}}

\usepackage{subcaption}
\usepackage{array}
\usepackage{caption}
\usepackage{booktabs}
\usepackage{multirow}
\usepackage[normalem]{ulem}
\hypersetup{colorlinks=true, linkcolor=blue, filecolor=magenta, urlcolor=cyan, citecolor=blue}

\makeatletter
\newcommand\fs@betterruled{%
  \def\@fs@cfont{\bfseries}\let\@fs@capt\floatl@ruled
  \def\@fs@pre{\vspace*{6pt}\hrule height.8pt depth0pt \kern2pt}%
  \def\@fs@post{\kern2pt\hrule\relax}%
  \def\@fs@mid{\kern2pt\hrule\kern2pt}%
  \let\@fs@iftopcapt\iftrue}
\floatstyle{betterruled}
\restylefloat{algorithm}
\makeatother

\renewcommand{\baselinestretch}{0.9705}

\begin{document}

\title{CII: Novel CSI-RS Metric for Joint Precoder and RIS Reporting in Multi-User NextG Networks}

 \author{
    \IEEEauthorblockN{Ali Fuat Sahin\IEEEauthorrefmark{1}\IEEEauthorrefmark{3},
    Sefa Kayraklik\IEEEauthorrefmark{1}\IEEEauthorrefmark{4}, 
    Ali Gorcin\IEEEauthorrefmark{1}\IEEEauthorrefmark{3}, 
    Ibrahim Hokelek\IEEEauthorrefmark{1}\IEEEauthorrefmark{3}, 
    Ertugrul Basar\IEEEauthorrefmark{5}\IEEEauthorrefmark{4},
    Halim Yanikomeroglu\IEEEauthorrefmark{7}}
    \IEEEauthorblockA{\IEEEauthorrefmark{1}Communications and Signal Processing Research (HISAR) Lab, TUBITAK BILGEM, Kocaeli, Turkiye}
    \IEEEauthorblockA{\IEEEauthorrefmark{3}Faculty of Electrical and Electronics Engineering, Istanbul Technical University, Istanbul, Turkiye}
    \IEEEauthorblockA{\IEEEauthorrefmark{4}Department of Electrical and Electronics Engineering, Koç University, Istanbul, Turkiye}
    \IEEEauthorblockA{\IEEEauthorrefmark{5}Department of Electrical Engineering, Tampere University, Tampere, Finland}
    \IEEEauthorblockA{\IEEEauthorrefmark{7}Non-Terrestrial Networks (NTN) Lab, Systems and Computer Engineering, Carleton University, Ottawa, ON, Canada}
    \IEEEauthorblockA{Email: \{ali.sahin, sefa.kayraklik, ibrahim.hokelek\}@tubitak.gov.tr, \\aligorcin@itu.edu.tr, ertugrul.basar@tuni.fi, halim@sce.carleton.ca} 

}   

\maketitle

\begin{abstract}
While reconfigurable intelligent surfaces (RISs) are among the key enablers for next-generation (NextG) wireless networks, efficient feedback reporting for joint base station (BS) precoding and passive RIS configuration remains a major challenge due to the associated signaling overhead. By extending the standard-compliant channel state information reference signal framework, this paper introduces a novel channel information indicator (CII) that jointly represents the active BS precoding matrix and passive RIS configuration within a single feedback metric for multi-user multiple-input single-output systems. Simulation results demonstrate that the proposed unified feedback framework significantly reduces uplink signaling overhead compared with conventional disjoint reporting schemes. Furthermore, despite only a modest increase in the feedback payload, the proposed CII-based scheme outperforms conventional precoding matrix indicator approaches in terms of system performance, offering a practical and standards-compatible solution for RIS integration in NextG wireless networks.
\end{abstract}


\section{Introduction} \label{sec:Introduction}
The evolution toward next-generation (NextG) wireless networks is driven by the demand for transformative applications and services with increasingly challenging requirements \cite{ITU-R_future}. According to the 3rd Generation Partnership Project (3GPP), sixth-generation (6G) systems are expected to support emerging use cases, including ubiquitous connectivity and computing, tactile internet, holographic communications, integrated sensing, and AI-native services, while simultaneously satisfying the stringent requirements of immersive, massive, hyper-reliable, and low-latency communications \cite{3gppTR22870, 11550739}. To this end, several enabling technologies are being incorporated into the radio access network, including integrated sensing and communication (ISAC), terahertz communications, non-terrestrial networks (NTNs), AI-native radio access networks, and reconfigurable intelligent surfaces (RISs) \cite{11456641}.

RISs have attracted significant attention as a key enabler of NextG wireless networks due to their ability to dynamically reconfigure the wireless propagation environment by adaptively controlling the phase and/or amplitude of incident electromagnetic waves \cite{8796365}. In practical multi-user multiple-input multiple-output (MU-MIMO) systems, network optimization, including active base station (BS) precoding and passive RIS phase-shift configuration, is typically performed at the BS \cite{11037049}. However, efficiently acquiring and reporting the channel and user equipment (UE) information required for the BS to jointly optimize these configurations remains a major challenge.

\subsection{Related Works} 

Feedback schemes for both precoding matrix and RIS configuration have been extensively studied in the literature; however, these two tasks have largely been addressed independently. For BS precoding, the authors in \cite{10319341} proposed a unified codebook design framework based on the channel state information reference signal (CSI-RS) feedback mechanism alongside MU-MIMO data transmission. The study in \cite{srsetal} proposed a sounding reference signal for CSI-RS, BS precoding, and feedback, using an end-to-end deep learning approach. For RIS configuration, the authors in \cite{11479675} investigated the impact of the RIS feedback overhead in terms of the number of transmitted bits. In \cite{orix2026etal}, an orchestration and control framework for the RIS was introduced for open radio access networks in smart wireless factories. Beyond independent optimization, several studies have considered the joint optimization of BS precoding and RIS configuration. For example, \cite{10542664} proposed a cross-layer framework for online BS precoding and RIS optimization to achieve fair scheduling in high-performance deployments. Similarly, in \cite{11555750}, a low-complexity singular-value-based method was proposed for RIS-assisted MIMO systems by jointly considering cascaded-channel RIS optimization and BS precoder selection.

Conventional cellular systems rely on standardized CSI-RS feedback to report precoding matrix indicator (PMI) for active BS beamforming, whereas passive RIS configurations are typically conveyed through separate quantization and compression schemes. Building upon these concepts, our prior work \cite{sahin2026novelcsirsreportingscheme} developed a CSI-RS reporting scheme to extract and report downlink complex channel information (CCI) for RIS optimization, demonstrating feasibility for RIS configuration but remaining limited to single-user scenarios. Consequently, a unified, standard-compatible feedback framework that jointly supports BS precoding and RIS configuration for multi-user networks remains an open research problem.

\subsection{Contributions}
This paper proposes a unified reporting framework for joint BS precoding and RIS configuration in multi-user NextG networks. By extending the standardized CSI-RS framework to multi-user, multiple-input single-output (MU-MISO) systems, a novel channel information indicator (CII) is introduced to jointly represent the BS precoding matrix and RIS configuration, thereby enabling coordinated beamforming and effective suppression of multi-user interference. The proposed reporting framework is designed to support multi-user channel feedback and multi-stream beamforming for single-antenna UEs. By consolidating the feedback required for BS precoding and RIS configuration into a unified reporting mechanism, the proposed approach significantly reduces uplink signaling overhead compared with conventional disjoint reporting schemes. Simulation results under practical scenarios demonstrate that the proposed CII-based framework consistently outperforms the conventional PMI-based approach with the same feedback budget, highlighting its potential for practical NextG physical layer implementations. Therefore, the proposed CII metric constitutes a promising approach for jointly reporting the BS precoding matrix and RIS configuration in RIS-assisted multi-user NextG wireless networks.

\section{System Model and Problem Formulation}
\label{sec:sysModel}
\begin{figure}[t] 
    \centering
    \includegraphics[width=0.95\columnwidth]{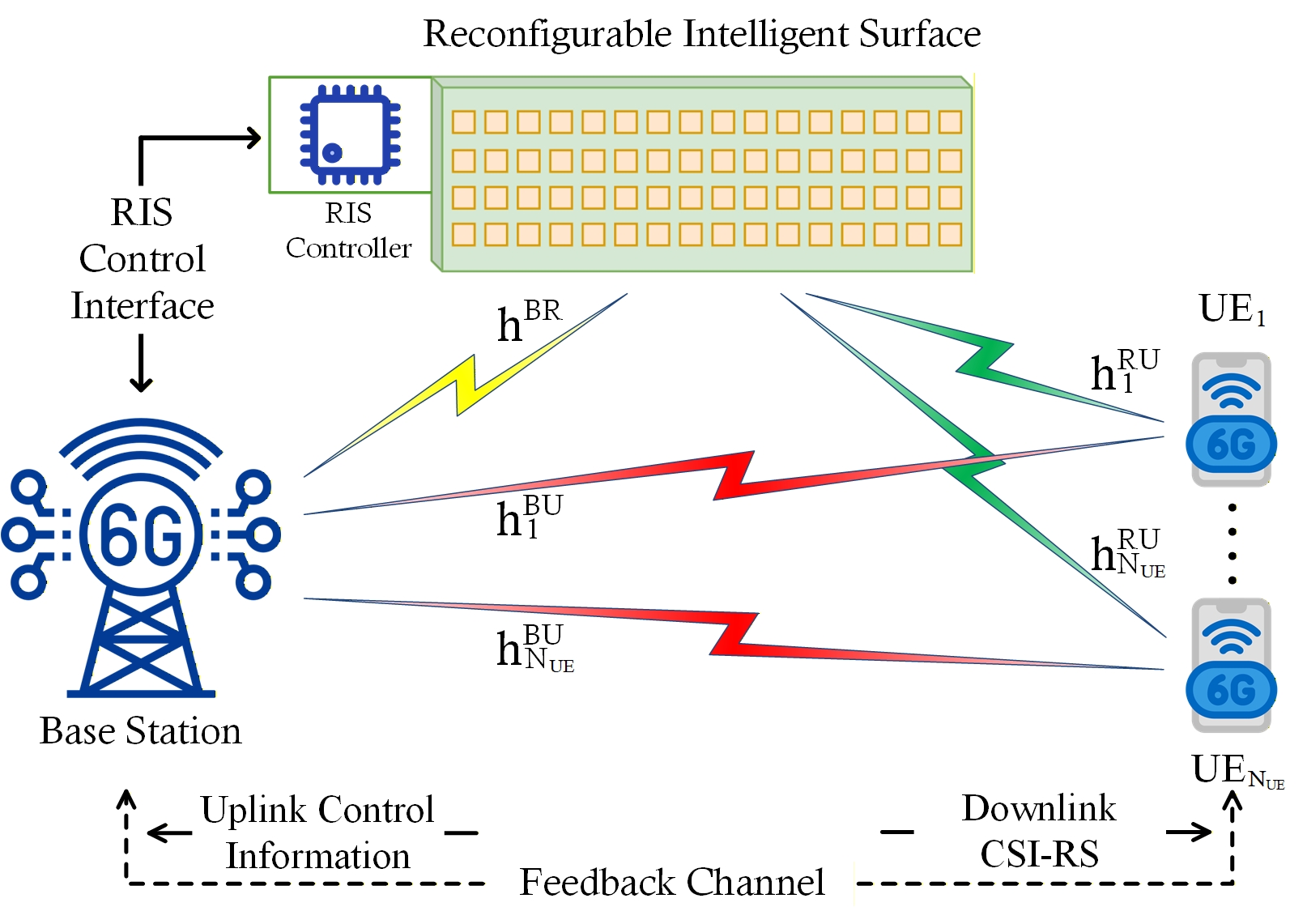}
    \caption{RIS-assisted MU-MISO system with feedback.}
    \label{fig:sysMod}
    \vspace{-15 pt}
\end{figure}

We consider a downlink (DL) RIS-assisted MU-MISO system, as illustrated in Fig. \ref{fig:sysMod}, where a BS equipped with $N_\mathrm{TA}$ transmit antennas serves $N_\mathrm{UE}$ single-antenna UEs with the assistance of a single RIS composed of $N_\mathrm{RIS}$ reflecting elements. The BS employs orthogonal frequency-division multiplexing (OFDM) with $N_\mathrm{SC}$ subcarriers spaced by $\Delta f$. Let $\mathbf{s}_{t,k,u} \in \mathbb{C}^{N_\mathrm{DS}\times 1}$ denote the DL data vector transmitted over subcarrier $k$ to the $u$-th UE at time index $t$, where $N_\mathrm{DS}$ is the number of data streams. Before transmission, the BS applies the precoding matrix $\mathbf{W}_{t,k} \in \mathbb{C}^{N_\mathrm{TA}\times N_\mathrm{DS}}$. Accordingly, the DL transmit signal is expressed as
\begin{equation}
    \mathbf{x}_{t,k,u} = \mathbf{W}_{t,k} \mathbf{s}_{t,k,u}.
    \label{eq:signalDefinition}
\end{equation}
Following, the end-to-end small-scale channel can be given as
\begin{equation}
    \mathbf{\tilde{h}}_{t,k,u} = \mathbf{\tilde{h}}^\mathrm{BU}_{t,k,u} + \mathbf{\tilde{h}}^\mathrm{RU}_{t,k,u} \mathbf{\Phi}_t \mathbf{\tilde{H}}^\mathrm{BR}_{t,k,u},
    \label{eq:smallScaleChannel}
\end{equation}
where $\mathbf{\tilde{h}}^\mathrm{BU}_{t,k,u} \in \mathbb{C}^{1 \times N_\mathrm{TA}}$ denotes the direct BS-UE channel, whereas $\mathbf{\tilde{h}}^\mathrm{RU}_{t,k,u} \in \mathbb{C}^{1 \times N_\mathrm{RIS}}$ and $\mathbf{\tilde{H}}^\mathrm{BR}_{t,k,u} \in \mathbb{C}^{N_\mathrm{RIS} \times N_\mathrm{TA}}$ denote the RIS-UE and BS-RIS channels, respectively.

Additionally, the large-scale effects for the BS-UE link can be modeled according to the 3GPP channel models \cite{etsiChannelModels} as
\begin{equation}
    PL^\mathrm{BU}_u(\mathcal{X}) = \mathcal{P}^\mathrm{BU}_\mathcal{X} \left(d^\mathrm{BU}_u, \lambda, h_\mathrm{BS}, h_\mathrm{UE}, I_{LoS}, \sigma_\mathrm{SF}\right),
    \label{eq:3gppPathLoss}
\end{equation}
where $\mathcal{X}$ is the scenario indicator w.r.t various deployment scenarios proposed in \cite{etsiChannelModels}, $d^\mathrm{BU}_u$ is the distance between BS and the $u$th UE, $\lambda$ is the wavelength of the transmitted signal, $I_{LoS}$ is the indicator of the LoS presence, $\sigma_\mathrm{SF}$ is the standard deviation of the shadow fading effects, $h_\mathrm{BS}$ and $h_\mathrm{UE}$ are the heights of the BS and the UE, respectively. Similarly, the cascaded path loss for a square RIS is expressed as \cite{etsiRisStandard}
\begin{equation}
    PL^\mathrm{BRU}_u = \frac{64 \pi^3}{G G_t G_r d_e^2 \lambda^2} \, {\left| \displaystyle \sum_{n=1}^{N_\mathrm{RIS}} \frac{\sqrt{F_n} \Gamma_n}{r^t_n r^r_n} e^{\frac{-j 2 \pi \left(r^t_n + r^r_n \right)}{\lambda}}\right|^{-2}},
    \label{eq:risPathLoss}
\end{equation}
where $G, G_t, G_r$ are the scattering gain for a single RIS element, transmit antenna gain, and receive antenna gain, respectively. Additionally, $d_e$ is the side length of a single RIS element. Moreover, $r^t_n$ and $r^r_n$ are the distances between the $n$th RIS element and the transmitter-receiver pair. Lastly, $\Gamma_n$ is the reflection coefficient of the $n$th RIS element, and $F_n$ is the combined power radiation of the transmitter, receiver, and RIS. Following, small-scale and large-scale effects can be combined to obtain the effective end-to-end channel, which can be expressed as
\begin{equation}
    \mathbf{h}_{t,k,u} = \sqrt{\frac{1}{PL^\mathrm{BU}_u}} \mathbf{\tilde{h}}^\mathrm{BU}_{t,k,u} + \sqrt{\frac{1}{\left. PL^\mathrm{BRU}_u \right|_{\Gamma_n=1}}} \mathbf{\tilde{h}}^\mathrm{RU}_{t,k,u} \mathbf{\Phi}_t \mathbf{\tilde{H}}^\mathrm{BR}_{t,k,u}.
    \label{eq:effectiveChannel}
\end{equation}
Additionally, $\mathbf{\Phi}$ denotes the RIS configuration under practical phase-dependent amplitude effects \cite{9115725}, and is expressed as
\begin{equation}
    \mathbf{\Phi} = \eta \;\mathrm{diag}\Big(\beta_1(\theta_1)e^{j\theta_1},  \ldots, \beta_{N_\mathrm{RIS}}(\theta_{N_\mathrm{RIS}}) e^{j\theta_{N_\mathrm{RIS}}}\Big),
    \label{eq:risConfiguration}
\end{equation}
where $\eta$ represents the RIS amplitude reflection efficiency,  $n$ denotes the RIS element index, and $\theta_n$ denotes the discrete phase shift which is expressed as
\begin{equation}
    \theta_n \in \mathcal{F} =\left\{ 0, \frac{2\pi}{\mathcal{L}}, \ldots, \frac{2 \pi \left(\mathcal{L}-1\right)}{\mathcal{L}} \right\} ,
    \label{eq:discretePhaseShifts}
\end{equation}
where $\mathcal{F}$ is the finite phase set and $\mathcal{L}$ represents the number of available RIS phase levels. Moreover, $\beta_n(\theta_n)$ represents the phase-dependent amplitude and can be shown as
\begin{equation}
    \beta_n(\theta_n) = \left( 1 - \rho_\mathrm{min} \right) \left( \frac{\mathrm{sin} \left( \theta_n - \phi \right) + 1}{2} \right)^\kappa + \rho_\mathrm{min},
    \label{eq:phaseDependentAmplitude}
\end{equation}
where $\rho_\mathrm{min} \geq 0$, $\phi \geq 0$, and $\kappa \geq 0$ are circuit-dependent parameters representing the minimum amplitude, horizontal shift, and steepness of the amplitude response curve, respectively. The received DL signal can then be expressed as
\begin{equation}
    y_{t,k,u} = \underbrace{\mathbf{h}_{t,k,u} \mathbf{x}_{t,k,u}}_{\text{Desired signal}} + \underbrace{\sum_{i=1, i \neq u}^{N_\mathrm{UE}} \mathbf{h}_{t,k,i} \mathbf{x}_{t,k,i}}_{\text{Interference}} + \underbrace{\mathbf{n}_{t,k,u}}_{\text{Noise}}
    \label{eq:receivedSignal}
\end{equation}
where $y_{t,k,u}$ denotes the received signal, which consists of the desired signal, interference, and additive noise components. Moreover, $\mathbf{n}_{t,k,u} \sim \mathcal{CN}(0,\sigma^2 \mathbf{I})$ denotes the complex Gaussian additive noise, where $\mathbf{I}$ is the identity matrix. Accordingly, the signal-to-interference-plus-noise ratio (SINR) is given by
\begin{equation}
    \gamma_{t,k,u} = \frac{\left| \mathbf{h}_{t,k,u} \mathbf{x}_{t,k,u} \right|^2}{\sum_{i=1, i \neq u}^{N_\mathrm{UE}} \left|\mathbf{h}_{t,k,i} \mathbf{x}_{t,k,i}\right|^2 + \sigma^2}.
    \label{eq:sinrFormula}
\end{equation}

Hence, the average sum spectral efficiency can be expressed as
\begin{equation}
    \mathcal{R} = \frac{1}{N_\mathrm{TD} N_\mathrm{SC}} \sum_{t=1}^{N_\mathrm{TD}} \sum_{k=1}^{N_\mathrm{SC}} \sum_{u=1}^{N_\mathrm{UE}} \mathrm{log}_2 \left(1 + \gamma_{t,k,u} \right), 
    \label{eq:averageSumRate}
\end{equation}
where $N_\mathrm{TD}$ represents the total number of OFDM symbols within the evaluation interval. Since the effective channel in \eqref{eq:effectiveChannel} and the resulting SINR in \eqref{eq:sinrFormula} depend jointly on the BS precoder and the RIS configuration, the system performance can be improved by jointly optimizing $\mathbf{W}_{t,k}$ and $\mathbf{\Phi}_t$. Accordingly, the spectral-efficiency optimization problem is formulated as
\begin{subequations}
    \begin{align}
        \max_{\mathbf{W}_{t,k}, \mathbf{\Phi}_t} \quad
        & \mathcal{R}\left(\mathbf{W}_{t,k}, \mathbf{\Phi}_t\right) \\
        \mathrm{s.t.} \quad
        & \sum_{k=1}^{N_\mathrm{SC}} \left\| \mathbf{W}_{t,k} \right\|_F^2
        \leq P_\mathrm{T}, \quad \forall t, \\
        & \theta_{n} \in \mathcal{F}, \quad \forall  n=1,2,\ldots,N_\mathrm{RIS},
    \end{align}
    \label{eq:problemFormulation}
\end{subequations}\hspace{-1em}
where $\left\|\cdot\right\|_F$ is the Frobenius norm and $P_\mathrm{T}$ is the maximum transmit power at the BS.

\section{Codebook Generation Methods}
\label{sec:codebook}
The optimization problem in \eqref{eq:problemFormulation} is non-convex due to the coupled dependence of the effective channel on the BS precoder $\mathbf{W}_{t,k}$ and the RIS configuration $\mathbf{\Phi}_t$, as well as the discrete phase constraints imposed by practical RIS hardware. Iterative and learning-based optimization methods can solve such problems. However, their computational complexity and convergence time may become prohibitive in dynamic wireless environments, where the channel coherence time and network delay budget impose challenging limits on the time available for such computations. To address this limitation, we employ a codebook-based formulation in which feasible BS precoders and RIS configurations are generated offline. These generated codebooks serve as the finite candidate sets for the compact CSI-RS-based reporting mechanism introduced in Section \ref{sec:framework}.

\subsection{Precoding Matrix Codebook Generation}
The PM codebook is constructed through a Type-II-inspired multi-user beam generation procedure based on an oversampled discrete Fourier transform (DFT) dictionary \cite{pmTutorialetal}. Let $N_\mathrm{PM}$ denote the number of PM codewords, and let $p \in \{1,\ldots,N_\mathrm{PM}\}$ be the PM codeword index. We first define an oversampled angular grid with $N_\mathrm{grid}=O_\mathrm{PM}N_\mathrm{TA}$ points, where $O_\mathrm{PM}$ is the oversampling factor. The corresponding DFT steering dictionary $\mathbf{A}_\mathrm{DFT}\in\mathbb{C}^{N_\mathrm{TA}\times N_\mathrm{grid}}$ is given by
\begin{equation}
    \mathbf{A}_\mathrm{DFT} = \left[\mathbf{a}_1,\mathbf{a}_2,\ldots,\mathbf{a}_{N_\mathrm{grid}}\right],
\end{equation}
where the $g$-th steering vector is expressed as
\begin{equation}
    \mathbf{a}_g = \frac{1}{\sqrt{N_\mathrm{TA}}} \left[1, e^{j2\pi\frac{g-1}{N_\mathrm{grid}}}, \ldots, e^{j2\pi(N_\mathrm{TA}-1)\frac{g-1}{N_\mathrm{grid}}}\right]^T,
\end{equation}
for $g\in\{1,\ldots,N_\mathrm{grid}\}$, and $(\cdot)^T$ denotes the transpose operation. For each codeword $p$ and UE $u$, a UE-specific local beam subset is formed by selecting $L_\mathrm{PM}$ adjacent steering vectors from the oversampled angular grid as
\begin{equation}
    \mathbf{B}_{p,u} = \left[\mathbf{a}_{g_{p,u,1}}, \mathbf{a}_{g_{p,u,2}}, \ldots, \mathbf{a}_{g_{p,u,L_\mathrm{PM}}} \right],
\end{equation}
where $\mathbf{B}_{p,u}\in\mathbb{C}^{N_\mathrm{TA}\times L_\mathrm{PM}}$, and $g_{p,u,\ell}$ denotes the $\ell$-th selected angular-grid index for the $u$-th UE under the $p$-th PM codeword, with $\ell=1,\ldots,L_\mathrm{PM}$. The selected beams are combined using the phase-only coefficient vector $\mathbf{c}^{(p)}=[c^{(p)}_1,\ldots,c^{(p)}_{L_\mathrm{PM}}]^T$, whose entries are given by
\begin{equation}
    c^{(p)}_\ell = \frac{1}{\sqrt{L_\mathrm{PM}}} e^{j2\pi\frac{(p-1)(\ell-1)}{N_\mathrm{PM}}}.
\end{equation}
Since $\|\mathbf{c}^{(p)}\|_2=1$, this construction yields a normalized phase-only combination of the selected DFT beams. Consequently, the assigned precoding vector is obtained as
\begin{equation}
    \mathbf{w}^{(p)}_u = \mathbf{B}_{p,u}\mathbf{c}^{(p)}.
\end{equation}
Then, the multi-user PM codeword is generated by stacking the UE-specific precoding vectors as
\begin{equation}
    \mathbf{W}^{(p)} = \left[\mathbf{w}^{(p)}_1, \mathbf{w}^{(p)}_2, \ldots, \mathbf{w}^{(p)}_{N_\mathrm{UE}}\right].
\end{equation}
Finally, each codeword is normalized as
\begin{equation}
    \mathbf{W}^{(p)} \leftarrow \frac{\mathbf{W}^{(p)}}{\left\|\mathbf{W}^{(p)}\right\|_F},
\end{equation}
and the resulting PM codebook is expressed as
\begin{equation}
    \mathcal{W} = \left\{\mathbf{W}^{(1)}, \mathbf{W}^{(2)}, \ldots, \mathbf{W}^{(N_\mathrm{PM})} \right\}.
    \label{eq:pmiCodebook}
\end{equation}

\subsection{RIS Configuration Codebook Generation}
The RIS configuration (RC) codebook is generated offline through a deterministic two-stage procedure. First, the RIS far-field angular space is divided into $N_\mathrm{RC}$ regions, each represented by a patch of adjacent far-field beams constructed using \cite{risFarFieldCodebook}. Then, a majorization-minimization (MM)-based multi-beam design adopted from \cite{risMultiBeamDesign} synthesizes a practical RIS coefficient vector for multi-UE operation. Let $r \in \{1,\ldots,N_\mathrm{RC}\}$ denote the RC codeword index, and factorize the angular grid as $N_\mathrm{RC}=G_yG_z$, where $G_y$ and $G_z$ are the numbers of region centers along the $y$- and $z$-axes, respectively. For the far-field stage, the steering vector along axis $\nu\in\{y,z\}$ is defined as
\begin{equation}
    \mathbf{u}_\nu(\alpha_\nu) = \left[1,e^{-j\pi\alpha_\nu},\ldots, e^{-j\pi(N_\nu-1)\alpha_\nu}\right]^T,
    \label{eq:steeringVector}
\end{equation}
where $N_\nu$ is the number of RIS elements along the $\nu$-axis and $\alpha_\nu \in [-1,1)$ is the directional cosine. The region centers are placed on a uniform grid as
\begin{equation}
    \bar{\alpha}_\nu(i_\nu) = -1+\frac{2i_\nu-1}{G_\nu}, \quad i_\nu = 1,\ldots,G_\nu,
\label{eq:regionCenters}
\end{equation}
and the two-dimensional far-field RIS beam pointing toward an arbitrary direction pair $(\alpha_y,\alpha_z)$ is given by
\begin{equation}
    \mathbf{q}(\alpha_y,\alpha_z) = \left[\mathbf{u}_y\!\left(\alpha_y\right) \otimes \mathbf{u}_z\!\left(\alpha_z\right)\right]^*,
    \label{eq:farFieldBeam}
\end{equation}
where $(\cdot)^*$ denotes conjugation and $\otimes$ denotes the Kronecker product. Each angular region is then covered by an $L_\mathrm{RC} \times L_\mathrm{RC}$ patch of adjacent beams. Denoting the region spacing along axis $\nu$ by $\nabla\nu = 2/G_\nu$, the intra-region beam offsets are defined as
\begin{equation}
    \delta_\nu(\iota) = \left(\iota - \frac{L_\mathrm{RC}+1}{2}\right)\frac{\nabla\nu}{ L_\mathrm{RC}}, 
\label{eq:patchOffsets}
\end{equation}
where $\iota = 1,\ldots,L_\mathrm{RC}$. The RC index $r$ is mapped to the corresponding angular-region indices as
\begin{equation}
    i_y = \left\lfloor \frac{\Pi(r-1)}{G_z} \right\rfloor + 1,
    \quad
    i_z = \big(\Pi(r-1) \bmod G_z\big) + 1,
    \label{eq:regionIndexMap}
\end{equation}
where $\Pi(\cdot)$ denotes the bit-reversal permutation of $\{0,\ldots,N_\mathrm{RC}-1\}$, which promotes a more uniform coverage of the angular space instead of concentrating consecutive codewords in neighboring regions. The $L_\mathrm{RC}^2$ patch beams associated with the selected region are stacked as
\begin{equation}
    \mathbf{\Xi}^{(r)} = \frac{1}{\sqrt{N_\mathrm{RIS}}}
    \Big[\, \mathbf{q}\big(\bar{\alpha}_y(i_y)+\delta_y(\iota),\;\bar{\alpha}_z(i_z)+\delta_z(\iota')\big) \,\Big]_{\iota,\iota'=1}^{L_\mathrm{RC}},
    \label{eq:patchMatrix}
\end{equation}
where $\mathbf{\Xi}^{(r)} \in \mathbb{C}^{N_\mathrm{RIS}\times L_\mathrm{RC}^2}$, the normalization term ensures unit-norm columns, and $\iota$ and $\iota'$ denote the patch-beam indices along the $y$- and $z$-axes, respectively. Given $\mathbf{\Xi}^{(r)}$, the multi-beam design problem is formulated as
\begin{equation}
    \min_{\boldsymbol{\theta}} \left\|\bar{\mathbf{g}} - \left(\mathbf{\Xi}^{(r)}\right)^H\boldsymbol{\theta}\right\|_2^2 \quad \mathrm{s.t.} \quad |\theta_n| = \beta_n(\theta_n) \leq 1,
    \label{eq:multiBeamDesignProblem}
\end{equation}
where $n = 1, \ldots, N_\mathrm{RIS}$, and $\bar{\mathbf{g}}$ and $\boldsymbol{\theta}$ represent the multi-beam target response vector and the ideal RIS reflection vector, respectively. To solve \eqref{eq:multiBeamDesignProblem}, the MM update step is given by
\begin{equation}
    \boldsymbol{\theta}^{(m+1)} = e^{\,j\angle\left( \mathbf{\Xi}^{(r)}\bar{\mathbf{g}}^{(m)} - \mathbf{\Xi}^{(r)} \left(\mathbf{\Xi}^{(r)}\right)^H \boldsymbol{\theta}^{(m)} + \lambda_{\max}\boldsymbol{\theta}^{(m)} \right)},
    \label{eq:updateStep}
\end{equation}
where $\lambda_{\max}$ is the maximum eigenvalue of $\left(\mathbf{\Xi}^{(r)}\right)^H\mathbf{\Xi}^{(r)}$. At iteration $m$, the target vector is updated as
\begin{equation}
    \bar{\mathbf{g}}^{(m)} = \sqrt{\frac{N_\mathrm{RIS}}{L_\mathrm{RC}^2}} \, e^{j\angle\left((\mathbf{\Xi}^{(r)})^H\boldsymbol{\theta}^{(m)}\right)},
    \label{eq:targetVector}
\end{equation}
which assigns equal target amplitudes to all $L_\mathrm{RC}^2$ beams in the patch. The update in \eqref{eq:updateStep} is repeated until convergence. After convergence, the continuous RIS phase vector is obtained as $\boldsymbol{\varphi}^{(r)}=\angle\boldsymbol{\theta}^{\star}$. The practical RIS coefficient $\phi_n^{(r)}$ is then computed by applying the discrete phase and phase-dependent amplitude model in \eqref{eq:discretePhaseShifts} and \eqref{eq:phaseDependentAmplitude}. Accordingly, the $r$-th RC codeword is constructed as
\begin{equation}
    \mathbf{\Phi}^{(r)} = \operatorname{diag} \left(\phi_1^{(r)}, \phi_2^{(r)}, \ldots, \phi_{N_\mathrm{RIS}}^{(r)}\right).
    \label{eq:rcCodeword}
\end{equation}
Repeating this procedure over all $N_\mathrm{RC}$ angular regions yields the RC codebook
\begin{equation}
    \mathcal{P} = \left\{\mathbf{\Phi}^{(1)},\mathbf{\Phi}^{(2)},\ldots,\mathbf{\Phi}^{(N_\mathrm{RC})}\right\}.
    \label{eq:rciCodebook}
\end{equation}

\section{CII-Based Reporting Framework}
\label{sec:framework}
To enable codebook-based optimization of \eqref{eq:problemFormulation}, we introduce a new feedback metric, referred to as the CII. Motivated by existing PMI-based feedback and RIS configuration reporting mechanisms \cite{pmTutorialetal, sahin2026novelcsirsreportingscheme}, the proposed CII jointly indexes a BS precoder and an RIS configuration. Hence, a single feedback index enables coordinated selection while reducing the reporting overhead compared with separate PMI and RIS configuration feedback. In a practical CSI-RS reporting scheme, the BS can configure a finite set of CII candidates, and the UE can report the preferred CII index after evaluating the CSI-RS resources. Thus, the proposed framework can be seen as a standards-oriented extension whose performance depends on practical factors such as feedback delay and CSI accuracy.

\begin{figure}[t]
    \centering
    \begin{subfigure}[b]{\columnwidth}
        \centering
        \includegraphics[width=\textwidth]{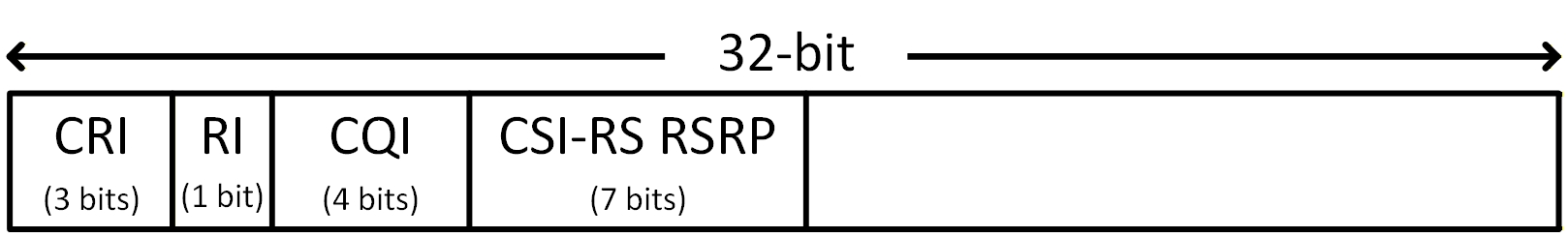}
        \caption{SISO case.}
        \label{subfig:sisoCase}
    \end{subfigure}
    \vfill
    \begin{subfigure}[b]{\columnwidth}
        \centering
        \includegraphics[width=\textwidth]{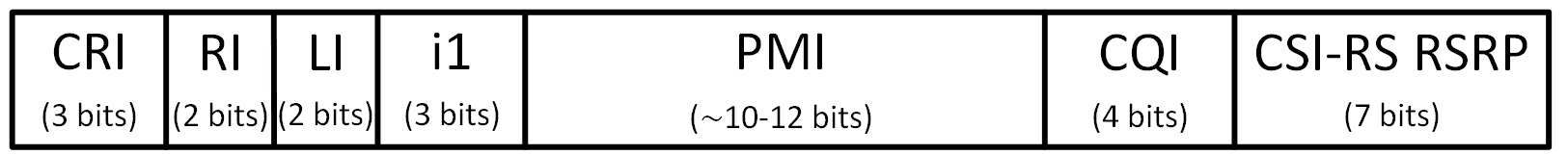}
        \caption{MIMO case.}
        \label{subfig:mimoCase}
    \end{subfigure}
    \vfill
    \begin{subfigure}[b]{\columnwidth}
        \centering
        \includegraphics[width=\textwidth]{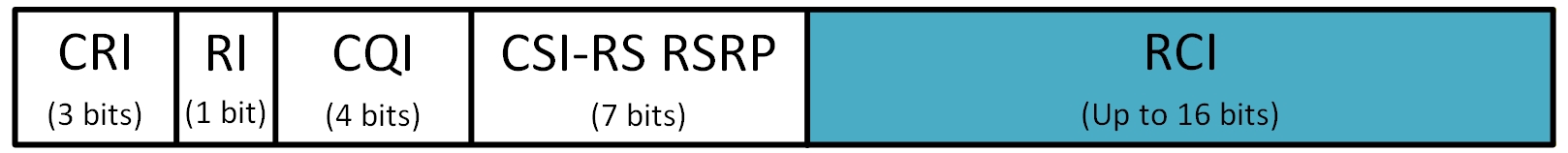}
        \caption{RIS case \cite{sahin2026novelcsirsreportingscheme}.}
        \label{subfig:risOnlyCase}
    \end{subfigure}
    \hfill
    \begin{subfigure}[b]{\columnwidth}
        \centering
        \includegraphics[width=\textwidth]{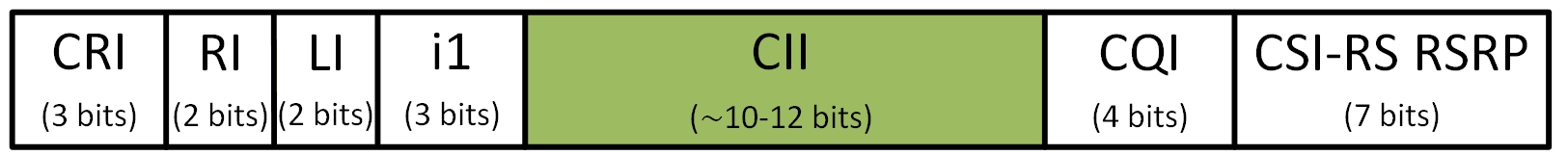}
        \caption{Proposed MIMO and RIS Case.}
        \label{subfig:ciiCase}
    \end{subfigure}
    \vspace{-10 pt}
    \caption{CSI-RS bitmaps for feedback length of 32 bits.}
    \label{fig:bitMap}
    \vspace{-10 pt}
\end{figure}

\subsection{PMI-Based Feedback}
As a conventional codebook-based CSI feedback mechanism, PMI is used in 5G NR to report the index of a preferred BS precoder selected from a finite codebook \cite{pmTutorialetal}. In this work, the PM codebook is given in \eqref{eq:pmiCodebook}, with $N_\mathrm{PM}=2^{B_\mathrm{PMI}}$, where $B_\mathrm{PMI}$ denotes the number of PMI feedback bits. Although PMI enables BS-side precoder adaptation, it does not carry any RIS configuration information. Therefore, in the considered network, PMI-only feedback cannot adapt the RIS reflection pattern to the channel conditions. Hence, we consider a PMI-only reporting case where the RIS is kept at a fixed reference configuration, denoted by $\mathbf{\Phi}^{(0)}$. This configuration can represent a mirror-like or predefined default RIS state. The reported PMI is then selected as
\begin{equation}
    p_\mathrm{PMI}^{\star} = \underset{p \in \{1,\ldots,2^{B_\mathrm{PMI}}\}}{\arg\max}\; \mathcal{R}\left(\mathbf{W}^{(p)}, \mathbf{\Phi}^{(0)}\right).
    \label{eq:pmiSelection}
\end{equation}
Fig. \ref{subfig:mimoCase} illustrates an exemplary 32-bit CSI-RS bitmap for PMI-only reporting in conventional MIMO applications.

\subsection{RCI-Based Feedback}
Similar to PMI-based precoder selection, RCI-based feedback reports the index of a preferred RIS configuration selected from a finite RC codebook. A related CSI-RS-based RIS reporting framework was considered in \cite{sahin2026novelcsirsreportingscheme} using CCI, whereas the RCI-only baseline here reports a compact RC codeword index. The RC codebook is given in \eqref{eq:rciCodebook}, with $N_\mathrm{RC}=2^{B_\mathrm{RCI}}$, where $B_\mathrm{RCI}$ denotes the RCI feedback length. To provide a complementary baseline, we consider an RCI-only reporting case in which only the RIS configuration is adapted, while the BS employs a fixed reference precoder denoted by $\mathbf{W}^{(0)}$. The reported RCI is then selected as
\begin{equation}
    r_\mathrm{RCI}^{\star} = \underset{r \in \{1,\ldots,2^{B_\mathrm{RCI}}\}}{\arg\max} \; \mathcal{R}\left( \mathbf{W}^{(0)}, \mathbf{\Phi}^{(r)} \right).
    \label{eq:rciSelection}
\end{equation}
Fig. \ref{subfig:risOnlyCase} illustrates an exemplary 32-bit CSI-RS bitmap for RCI-only reporting.

\subsection{CII-Based Feedback}
To reduce feedback overhead while enabling coordinated selection of the BS precoder and RIS configuration, we introduce CII-based feedback. Unlike PMI-only or RCI-only feedback, the CII directly indexes a paired BS precoder and RIS configuration. Accordingly, the CII codebook is defined as
\begin{equation}
    \mathcal{C} = \left\{\left(\mathbf{W}^{(1)},\mathbf{\Phi}^{(1)}\right),\ldots,\left(\mathbf{W}^{\left(2^{B_\mathrm{CII}}\right)},\mathbf{\Phi}^{\left(2^{B_\mathrm{CII}}\right)}\right)\right\},
    \label{eq:ciiCodebook}
\end{equation}
where $B_\mathrm{CII}$ denotes the number of CII feedback bits. In this work, we set $B_\mathrm{CII}=\max\left(B_\mathrm{PMI},B_\mathrm{RCI}\right)$ to provide a single-index feedback mechanism whose feedback length is comparable to previous reporting methods. The reported CII is then selected by maximizing the spectral efficiency as
\begin{equation}
    q_\mathrm{CII}^{\star} = \underset{q \in \{1,\ldots,2^{B_\mathrm{CII}}\}}{\arg\max}\; \mathcal{R}\left(\mathbf{W}^{(q)},\mathbf{\Phi}^{(q)}\right).
    \label{eq:ciiSelection}
\end{equation}
Thus, CII-based feedback enables coordinated precoder-RIS configuration reporting without requiring separate feedback indices. Fig. \ref{subfig:ciiCase} illustrates an exemplary 32-bit CSI-RS bitmap corresponding to the proposed CII-based reporting case.

\subsection{Separate PMI and RCI Feedback}
Although CII-based feedback reduces the reporting overhead by using a single paired index, its search space is limited to the predefined pairs in $\mathcal{C}$. As an upper-bound benchmark, we also consider separate PMI and RCI feedback, where the BS precoder and RIS configuration are jointly selected over the Cartesian product of their respective codebooks. The corresponding index pair is obtained as
\begin{equation}
    \left(p_\mathrm{PMI}^{\star}, r_\mathrm{RCI}^{\star}\right) = \underset{ \substack{ p \in \{1,\ldots,2^{B_\mathrm{PMI}}\} \\ r \in \{1,\ldots,2^{B_\mathrm{RCI}}\}}}{\arg\max} 
    \; \mathcal{R}\left(\mathbf{W}^{(p)}, \mathbf{\Phi}^{(r)}\right).
    \label{eq:separatePMIRCISelection}
\end{equation}
This approach explores a larger search space and may achieve higher spectral efficiency than CII-based feedback, but at the cost of two separate feedback indices. The corresponding feedback overhead is given by
\begin{equation}
    B_\mathrm{PMI+RCI} = \log_2\left|\mathcal{W} \times \mathcal{P}\right| >  \log_2\left|\mathcal{C}\right| = B_\mathrm{CII}.
\label{eq:feedbackOverheadComparison}
\end{equation}
Therefore, CII aims to approach the performance of separate PMI and RCI feedback while maintaining the overhead of a single-index reporting mechanism.

\begin{figure*}[t]
    \centering
    \begin{subfigure}{0.66\columnwidth}
        \centering
        \includegraphics[width=\columnwidth]{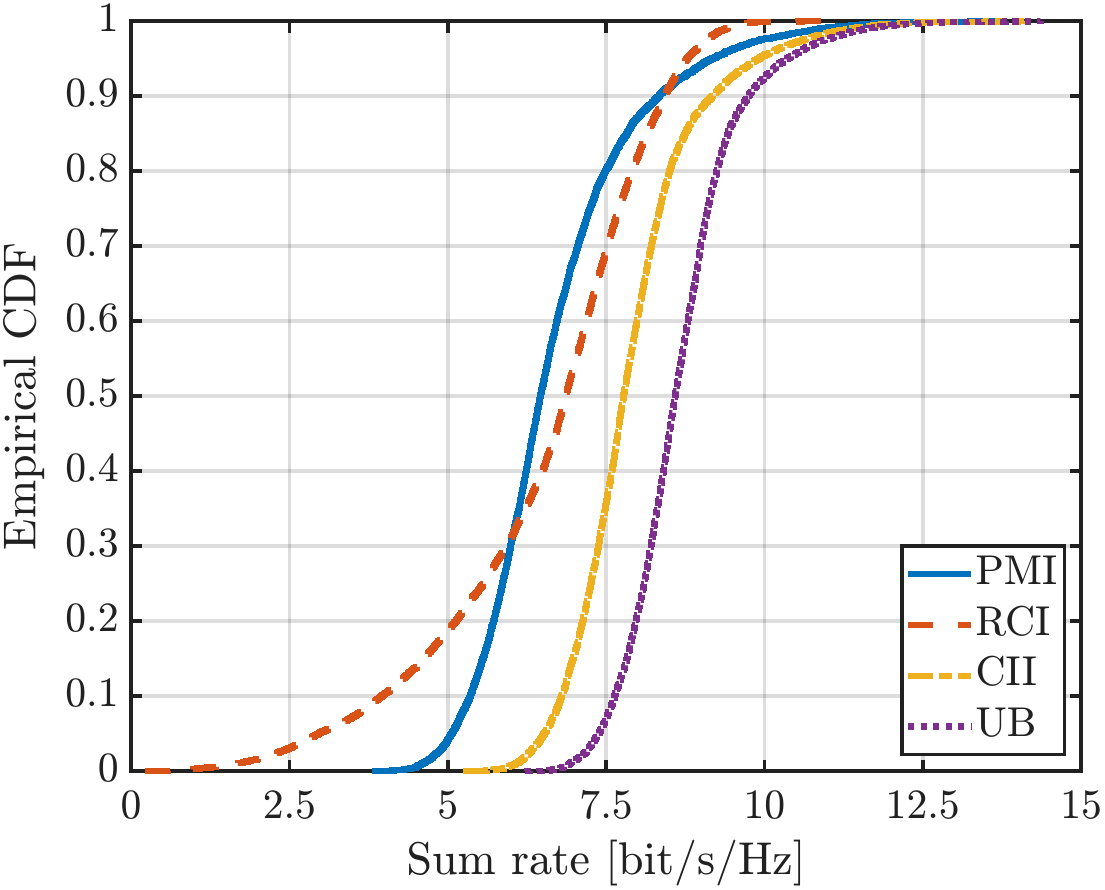}
        \caption{Empirical CDF for $B=12$.}
        \label{subfig:empiricalCDf}
    \end{subfigure}
    \hfill
    \begin{subfigure}{0.66\columnwidth}
        \centering
        \includegraphics[width=\columnwidth]{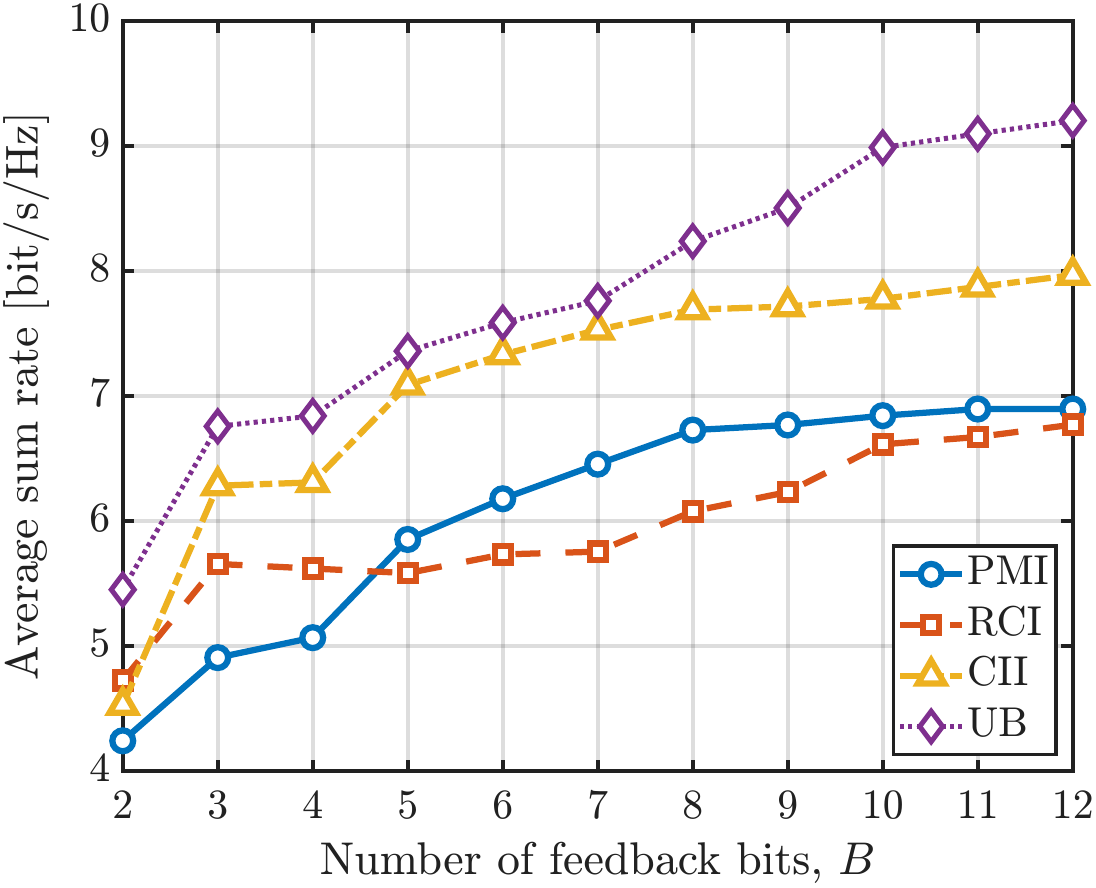}
        \caption{Spectral efficiency versus $B$.}
        \label{subfig:bitLength}
    \end{subfigure}
    \hfill
    \begin{subfigure}{0.66\columnwidth}
        \centering
        \includegraphics[width=\columnwidth]{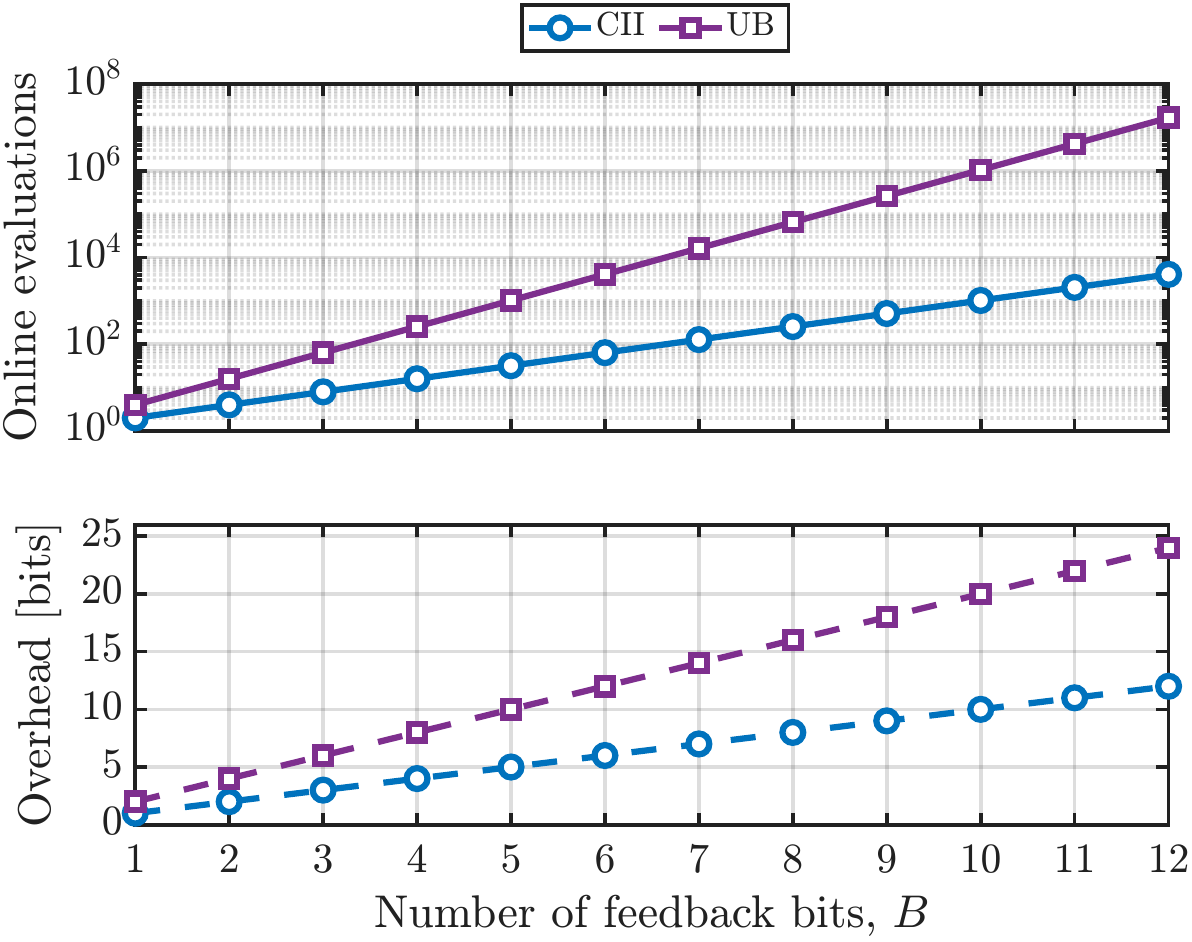}
        \caption{Evaluation and feedback overhead.}
        \label{subfig:complexity}
    \end{subfigure}
    \caption{Performance comparison of the proposed reporting framework.}
    \label{fig:simResults}
    \vspace{-10 pt}
\end{figure*}

\section{Simulation Scenarios and Results}
\label{sec:simResults}
In this section, simulation results are presented to evaluate the proposed CII-based reporting framework. We consider a three-dimensional RIS-assisted MU-MISO deployment, where the BS and RIS are located at $(0,0,10)$ m and $(25,25,5)$ m, respectively. The UEs are uniformly distributed within a circular service region of radius $100$ m at a height of $1.5$ m, while remaining in the RIS far-field region. The carrier frequency is set to $f_c=7.5$ GHz, with bandwidth $\mathcal{B}=200$ MHz and subcarrier spacing $\Delta f=60$ kHz, to emulate the FR3 frequency range \cite{11244718}. The BS-UE, BS-RIS, and RIS-UE links are modeled using CDL-D channels, while the large-scale fading of the BS-UE link follows the 3GPP UMi-Street Canyon model \cite{etsiChannelModels}. The cascaded RIS path loss follows the RIS-assisted propagation model in Section \ref{sec:sysModel}, with parameters adopted from \cite{11555750}, and the practical RIS parameters $(\rho,\phi,\kappa)$ are adopted from \cite{9115725}. The PM and RC codebooks contain $N_\mathrm{PM}=N_\mathrm{RC}=4096$ candidates ($12$-bit feedback budget). The main simulation parameters are summarized in Table \ref{tab:simParameters}.

\begin{table}[t]
\large
\centering
\caption{Simulation Parameters}
\vspace{-5 pt}
\label{tab:simParameters}
\resizebox{0.7\columnwidth}{!}{%
\begin{tabular}{c|c||c|c}
    \hline
    \hline
    {\textbf{Parameter}} & {\textbf{Value}} & {\textbf{Parameter}} & {\textbf{Value}}  \\
    \hline
    \multicolumn{4}{c}{BS Parameters} \\
    \hline
    $N_\mathrm{TA}$ & $32$ & $P_{\mathrm{T}}$ & $20$ dBm \\
    \hline
    $N_\mathrm{SC}$ & $3168$ & $N_\mathrm{TD}$ & $56$ (1 ms)  \\
    \hline
     $G_t$ & $10$ dBi & $\sigma_\mathrm{SF}$ & $4$ dB \\
    \hline
    $L_\mathrm{PM}$ & 4 & $O_{\mathrm{PM}}$ & 4 \\
    \hline
    \multicolumn{4}{c}{RIS Parameters} \\
    \hline
    $N_\mathrm{RIS}$ & $1024$ & $d_e$ & $2$ cm \\
    \hline
    $\mathcal{L}$ & 4 & $L_\mathrm{RC}$ & $2$ \\
    \hline
    $F$ & 1 & $G$ & 1 dBi \\
    \hline
    \multicolumn{4}{c}{UE Parameters} \\
    \hline
    $N_\mathrm{UE}$ & $4$ & $G_r$ & $3$ dBi \\
    \hline
    \hline
\end{tabular}}
\vspace{-13 pt}
\end{table}

The proposed reporting framework is evaluated by comparing CII with PMI, RCI, and separate PMI+RCI reporting. The results first examine the average sum-rate distribution under a fixed feedback budget, followed by the impact of feedback bit length and reporting cost. Fig. \ref{subfig:empiricalCDf} illustrates the empirical cumulative distribution function (CDF) of the average sum rate in \eqref{eq:averageSumRate}. The CII-based scheme shifts the distribution to the right compared with the PMI and RCI baselines, confirming the performance advantage of jointly indexing the PM and RC codebooks. RCI reporting exhibits a wider low-rate tail, indicating that sole RIS adaptation may be less robust when the BS precoder remains fixed. The separate PMI and RCI benchmark achieves the best performance by independently selecting the PM and RC indices, acting as an upper bound. The remaining gap between CII and this upper bound shows the performance loss caused by using a single-index feedback.

The average sum-rate performance as a function of the feedback bit length $B$ is shown in Fig. \ref{subfig:bitLength}. The overall trend shows that larger feedback bits improve the achievable rates by providing finer codebook resolution for PMI, RCI, and CII selection. CII generally provides higher average sum rates than both baselines, confirming the benefit of jointly indexing the PM and RC codebooks. As shown in Fig. \ref{subfig:bitLength}, CII requires only $5$ bits to achieve the same converged average sum rate that both baselines attain with $12$ bits. The separate PMI and RCI benchmark achieves the highest rate by searching over independently selected PM and RC indices; however, this performance gain comes at a significant cost, as shown in Fig. \ref{subfig:complexity}.  Specifically, PMI, RCI, and CII each require $\mathcal{O}(2^B)$ rate evaluations and a feedback payload of $B$ bits, whereas separate PMI and RCI feedback increases the search complexity to $\mathcal{O}(2^{2B})$ and requires $2B$ feedback bits. Hence, CII offers a favorable performance-complexity trade-off with single-index feedback cost while achieving higher sum-rate performance.

\section{Conclusion \& Future Works}
\label{sec:conc}

This paper addressed the challenge of channel feedback overhead in practical RIS-assisted NextG networks. By extending the standard-compliant CSI-RS framework, a unified feedback scheme was proposed through the introduction of a novel channel information indicator for jointly reporting BS precoding matrices and RIS configurations in MU-MISO systems. Simulation results demonstrate that the proposed framework achieves the performance of conventional baselines using only $5$ bits, significantly reducing uplink signaling overhead while effectively mitigating multi-user interference. Furthermore, despite only a modest increase in the feedback payload, the proposed framework outperforms the conventional PMI-based approach, demonstrating its potential for practical physical layer implementation. Therefore, the results indicate that the proposed CII metric is an effective solution for jointly reporting the BS precoding matrix and RIS configuration in RIS-assisted multi-user NextG wireless networks. Future work will focus on validating CII in operational testbeds, extending the framework to MIMO user deployments, and developing CII-based optimization schemes that jointly exploit active BS precoding and passive RIS configuration in NextG networks.

\balance
\bibliographystyle{IEEEtran}
\bibliography{refer}

\end{document}